\documentclass[pra,11pt,onecolumn]{revtex4}

\usepackage{amsmath,amssymb,graphicx,amsbsy}
\usepackage{calc}
\usepackage{numprint}
\usepackage{hyperref}
\usepackage[FIGTOPCAP,raggedright,nooneline,bf,footnotesize]{subfigure}
\usepackage{indentfirst}
\usepackage[usenames,dvipsnames]{pstricks}

\newcommand{\figref}[1]{Fig.~\ref{#1}}

\newcommand{\mathsym}[1]{{}}

\newcommand{\kp}[1]{{\pmb{k}}_{#1 \perp}}

\renewcommand{\eqref}[1]{Eq.~(\ref{#1})}
\newcommand{\tabref}[1]{Tab.~(\ref{#1})}
\renewcommand{\figref}[1]{Fig.~\ref{#1}}

\newcommand{\mc}[1]{\mathcal{#1}}

\begin{document}


\title{Towards correcting atmospheric beam wander via pump beam control in a down conversion process}
\author{Christopher J. Pugh,$^{1,2}$ Piotr Kolenderski,$^{1,2,3,*}$ Carmelo Scarcella,$^4$ Alberto Tosi$^4$ and Thomas Jennewein$^{1,2}$}
\affiliation{
$^1$Institute for Quantum Computing, University of Waterloo, 200 University Ave.~West, Waterloo, Ontario, Canada, N2L 3G1
\\
$^2$Department of Physics and Astronomy, University of Waterloo, 200 University Ave.~West, Waterloo, Ontario, Canada, N2L 3G1
\\
$^3$Faculty of Physics, Astronomy and Informatics, Nicolaus Copernicus University, Grudziadzka 5, 87-100 Torun, Poland
\\
$^4$Politecnico di Milano, Dipartimento di Elettronica, Informazione a Bioingegneria,
Piazza Leonardo da Vinci 32, I-20133 Milano, Italy\\
$^*$Corresponding author:  kolenderski@fizyka.umk.pl
}

\begin{abstract}
Correlated photon pairs produced by a spontaneous parametric down conversion (SPDC) process can be used for secure quantum communication over long distances including free space transmission over a link through turbulent atmosphere. We experimentally investigate the possibility to utilize the intrinsic strong correlation between the pump and output photon spatial modes to mitigate the negative targeting effects of atmospheric beam wander. Our approach is based on a demonstration observing the deflection of the beam on a spatially resolved array of single photon avalanche diodes (SPAD-array). 
\end{abstract}
\date{\today}

\maketitle
\bibliographystyle{ol}
\bibliography{AdaptSpatCorr}

\section{Introduction}

Single photons generated in the process of spontaneous parametric down conversion (SPDC) can provide information carriers suitable for secure quantum communication. Currently, there is a large field of interest in performing quantum communication over long distance free space links \cite{Schmitt07, Bourgoin2013, Bonato2009}. Quantum communication protocols only use the signals that are collected by the receiver and, therefore, the system is not as negatively affected by occasional link drop outs as classical communication protocols. In order to collect as many photons as possible at the receiver it is necessary to have a mechanism to guide the photons to the receiver. Using traditional adaptive optics \cite{Beckers1993,Lipinski1994}, the transmission beam can be manipulated in such a way that the effects of atmospheric turbulence can be mitigated. However, due to the sensitivity of the quantum degree of freedom of the photon used in free space transmission (in many cases polarization), these methods can easily alter the state and render the transmission useless. These techniques also add additional loss to the quantum transmission channel, which reduces the amount of photons the receiver can gather.   

Spontaneous parametric down conversion \cite{Mandel1987} has strong correlations between the directions of the pump and daughter photons due to the phase matching conditions. By manipulating the spatial properties of the pump beam, one can change the spatial modes of the down converted photons without affecting their polarization or other properties. Therefore, by manipulating the pump beam, potentially with tip-tilt mirrors or other means, it is possible to enhance a link quality by altering the directions of transmitted daughter photon without direct interaction in the quantum transmission channel. In a quantum communication scheme it is beneficial to manipulate the pump beam since the pump can have a narrow bandwidth (unlike the SPDC photons) and can handle transmission losses more easily than the SPDC photons. These correlations, which are an important feature for the quality of transmission in free space, have already been studied in the context of aberration cancellation \cite{Minozzi2013,Filpi2015}. Spatial correlations \cite{daCostaMoura2008} have already been investigated using a scanning \cite{Calderon-Losada2016} and scanning free space detector \cite{Zeilinger2005a,Neves2007,Lima2008,Lima2011} and intensified CCD cameras \cite{Fickler2013,Prabhakar2014}. 

To test the effect of spatial correlations between the pump and the SPDC photons we use an array of single photon avalanche diodes (SPAD) \cite{Zappa2007,Guerrieri2010a, Scarcella2013, Kolenderski2014, Johnsen2014}, which offer temporal and spatial resolution on a single photon level. Here we investigate the possibility of controlling the spatial characteristics of one of the photons produced from SPDC by altering the direction of the pump beam (ultimately to be modified by means of tip-tilt correction).

\section{Spatial correlations in SPDC} \label{sec:theory}
\label{sec:SpatSec}
At this point we would like to characterize the spatial correlations in SPDC. We have altered a common SPDC setup to allow for the pump entry direction into a nonlinear crystal to be manipulated. The resulting photons propagate in the free space and the signal photon is analyzed using a SPAD array and the idler photon is coupled into multimode fiber.

The first goal is to find an approximate analytical condition for the strength of spatial correlation in the parametric down conversion process. The probability amplitude of the signal (idler) photon propagating in a direction parametrized by $\kp{s}$ ($\kp{i}$), assuming the pump photon central propagation direction was $\kp{p}$, can be found using the framework given in Ref.~\cite{Kolenderski2009, Kolenderski2009a}:
\begin{multline}
    \Psi(\kp{s},\kp{i},\kp{p};\omega_s,\omega_i,\omega_p)= \\ \mc{N} \Lambda(\omega_i)    \exp\left(-\frac{w_p^2}{2}(\kp{s}+\kp{i}-\kp{p})^2 \right)   
     \text{sinc}\left(\frac{1}{2} L \Delta k_z(\kp{s},\kp{i}, \kp{p};\omega_s,\omega_i,\omega_p)  \right).
\end{multline}
Where $\mc{N}$ is a normalization factor, $\omega_s$ ,$\omega_i$, $\omega_p$ are signal, idler and pump angular frequencies,  $ \Lambda(\omega_i)$ represents a bandpass interference filter transmission amplitude, and the third component describes a pump beam spatial profile that we assumed to be Gaussian with the characteristic width $w_p$. The last term is a phase matching function, where a phase mismatch is defined as: $\Delta k_z(\kp{s},\kp{i},\kp{p};\omega_s,\omega_i,\omega_p)=k_{pz}(\kp{s}+\kp{i}-\kp{p},\omega_p)-k_{sz}(\kp{s},\omega_s)-k_{iz}(\kp{i},\omega_i)$ and $L$ is the length of the crystal.

In order to arrive at an analytical solution, we expand the phase mismatch, $\Delta k_z(\kp{s},\kp{i},\kp{p};\omega_s,\omega_i,\omega_p)$, up to first order in the signal and idler photon's transverse wavevectors around the directions $\kp{s0}$, $\kp{i0}$ and frequencies around $\omega_p/2$. The directions and frequency are determined by the perfect phase matching condition $\Delta k_z(\kp{s0},\kp{i0},\kp{p};\omega_p/2,\omega_p/2,\omega_p)=0$ and are related to our experimental setup as visualized in the inset in \figref{fig:Setup}. Note that the pump photon central direction $\kp{p}$ and its central frequency $\omega_p$ are fixed, therefore $\kp{p}=\kp{s0}+\kp{i0}$. The expansion yields:
\begin{equation}
  \Delta k_z(\kp{s},\kp{i},\kp{p};\omega_s,\omega_i,\omega_p)\approx  \mathbf{d}_{s}(\kp{s}-\kp{s0})+\mathbf{d}_{i}(\kp{i}-\kp{i0})+{\bf \beta_s}(\omega_i-\frac{1}{2}\omega_p)+{\bf \beta_i}(\omega_s-\frac{1}{2}\omega_p),
\end{equation}
where for the sake of compact notation we introduce two dimensional vectors \cite{Kolenderski2009}:
\begin{equation}
	\mathbf{d}_{\mu}=\frac{d}{d \kp{  \mu}}  \Delta k_{z}(\kp{s},\kp{i},\kp{p};\omega_s,\omega_i,\omega_p) , \quad \mu=s,i,
\end{equation}
and we denote the coefficients related to the dispersion as:
\begin{equation}
	\beta_{\mu}=\frac{d }{d \omega_{\mu}}  \Delta k_{z}(\kp{s},\kp{i},\kp{p};\omega_s,\omega_i,\omega_p), \quad \mu=s,i.
\end{equation}
The above approximation is legitimate as we are interested only in paraxial propagation and will use bandpass filters. We assumed a continuous wave pumping laser of fixed angular frequency $\omega_p$, which allows the parametrization of the wave function with only one frequency, $\omega_i$, chosen to be the idler photon since $\omega_p=\omega_s+\omega_i$. This expansion in conjunction with the Gaussian approximation of the sinc function given by:
$\mathop{\text{sinc}} x\approx \exp(- {x^2}/5 ),$ allows one to get the biphoton wave function in the following form: 
\begin{multline}\label{eq:wf:spat:approx}
	\Psi(\kp{s},\kp{i},\kp{p};\omega_i,\omega_p) \approx
	\mc{N}  \Lambda(\omega_i)   \exp\left(-\frac{w_p^2}{2}(\kp{s}+\kp{i}-\kp{p})^2\right)\times\\
\exp\left(-\frac{L^2}{10}\left(\mathbf{d}_{s}(\kp{s}-\kp{s0})+\mathbf{d}_{i}(\kp{i}-\kp{i0})\right.\right. +\left.\left.{\bf \beta}_s(\omega_i-\frac{1}{2}\omega_p)+{\bf \beta}_i(\frac{1}{2}\omega_p-\omega_i)\right)^2\right.\Bigg).
\end{multline}  

The aim is to characterize the correlation between the signal photon and the pump beam keeping the idler photon direction fixed. In this work we only consider a linear, one dimensional detector aligned along the $x$ axis, therefore we can drop $y$ coordinates in our analysis.  The scenario is depicted in \figref{fig:Setup}. We are interested in the coincidence probability distribution $P(k_{sx})$ of a signal photon detection for the transverse direction, $k_{sx}$, and the idler photon coupled into multimode fiber.  The fiber is placed along the direction of perfect phase matching in the case where the pump photon is at normal incidence to the nonlinear crystal, which is $\kp{p}=0$. We also take into account the finite size of our SPAD array detectors, which results in a certain range of directions $\delta k$ that each of the pixels monitor. The above observations allow us to write a formula for the probability of finding the signal photon in a direction $k_{sx}$ given the detection of the idler photon:
\begin{equation}
P(k_{sx})=\int_{k_{sx}-{\delta k}/{2}}^{k_{sx}+{\delta k}/{2}} d k_{s\perp} \int d k_{i\perp}\  d \omega_i\ |\Psi(\kp{s},\kp{i},\kp{p};\omega_i,\omega_p)|^2.
\label{eq:prob:SPAD}
\end{equation}
Using  \eqref{eq:wf:spat:approx} we evaluate the above probability function in analytical form. It is rather long so we don't present it here, it can be found in Ref.~\cite{Pugh2013}.

The next step is to obtain the relation between the position of the SPAD array detector, $x$, and the propagation direction parametrized by the transverse spatial direction $k_{sx}$. We assume that the SPAD array is placed at the position close to the perfect focal plane of the lens. Let's assume the signal photon travels toward the SPAD array and has a transverse spatial wave function, $\psi(x)$, at position $z$ propagating along the $z$ axis. The photon propagates in free space for a distance $p$, then through lens L3 with focal length $f$ and then is analyzed at a distance $q$ after the lens L3. The standard paraxial Fresnel propagation results in the following relation between the Fourier transform of the initial wave function, $\tilde\psi_{in}(k_x)$, and the final one, $\psi_{out}(x)$:
\begin{equation}
\psi_{out}(x) \propto \tilde\psi_{in}\left(\frac{2 \pi D}{\lambda p q}x\right),
\label{eq:propagation}
\end{equation}
where $1/D=1/f-1/p-1/q$, $\lambda$ is the wavelength of the propagating photon.

Let's consider the spatial wave function, which is a Gaussian centered around the direction $k_{sx0}$:
\begin{equation}
\tilde\psi(k_{sx})= N \exp\left(-\frac{1}{2}w_s^2(k_{sx}-k_{sx0})^2\right),
\end{equation}
where $2 w_s$ is the beam's diameter at $13.5\% $ of maximal intensity, $k_{sx0}$ the x component of central propagation direction and $N$ is a normalization factor. After the propagation as in \eqref{eq:propagation} one can get the following relations for the position of the maximum, $x_0$, and the beam radius, $w_{out}$:
\begin{eqnarray}
x_0 = k_{sx0} \frac{\lambda p q}{2 \pi D}, \\ 
w_{out} = \frac{1}{w_s} \frac{\lambda p q}{2 \pi D}. 
\label{eq:scaling}
\end{eqnarray}
The first equation gives a mapping between the central propagation direction of the Gaussian beam and actual position of the maximum of the Gaussian profile measured by the SPAD array. The last equation gives the relations of the beam inside the nonlinear crystal and the width measured by the SPAD array. Assuming the SPAD array is placed perfectly in the focal plane of the lens, meaning $p=f$, it is easy to find that the ratio of positions of the maxima $x_0/x^{(f)}_0$ is inversely proportional to the ratio of beam diameters $w_{out}/w^{(f)}_{out}$. The superscript $f$ here denotes the case of perfect alignment. This relation can be used to compare our numerical model with the experimental results.

Taking into account the parameters of our experimental setup, using our model \eqref{eq:prob:SPAD} we expect the characteristic width of the signal photon to be  $w_s=1432(3) \mu$m. Moreover the central direction $\alpha_{s0}$ of the signal photon follows a linear model given by $\alpha_{s0}=3.00000(9)+2.0109(14)\alpha_p$, where $\alpha_{s0}$ and $\alpha_p$ are measured in degrees. 

\section{Experiment} \label{sec:experiment}
\begin{figure}
\centering
\includegraphics[width=0.65\columnwidth,angle=0]{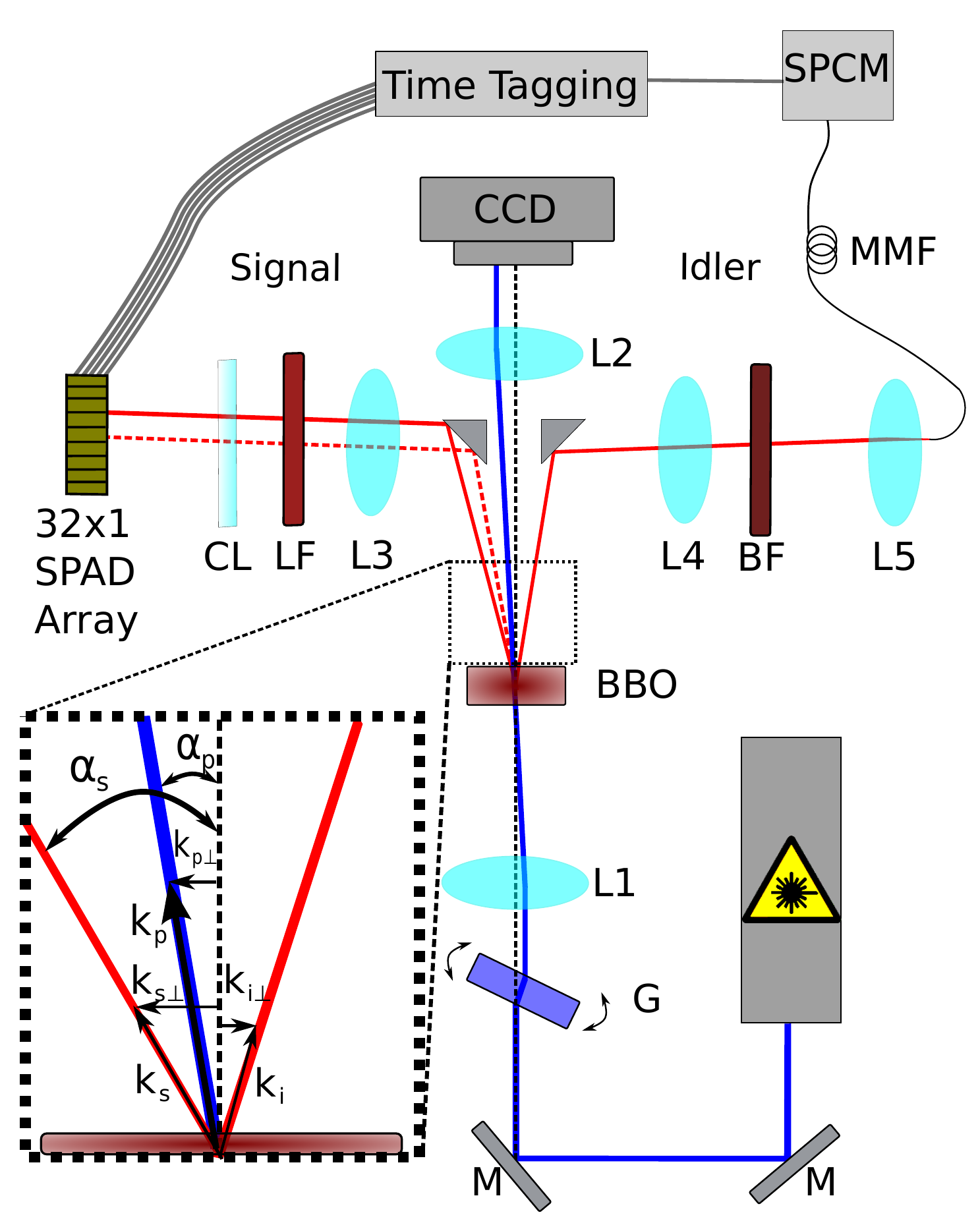}
\caption{Experimental Setup. A laser at 404 nm is directed by a mirror through a piece of NBK7 glass G which can turn to alter the input beam angle to the BBO crystal. The first lens L1 ($f= 30$ cm) focuses the beam at the crystal and the next lens L2 ($f= 50$ cm) collimates the beam for a position measurement on the CCD camera. The idler photon spatial mode is collimated by lens L4 ($f= 30$ cm) and passes through a bandpass filter  (Thorlabs FB810-10, 810 nm, FWHM 10 nm), and is coupled by lens L5 into a multimode fiber (SMF-28e+ supporting 7 modes at $810$ nm). Finally it is detected by a SPCM (PerkinElmer, SPCM-AQ4C). The signal beam is collimated by the lens L3 ($f = 30$ cm) and passes through a longpass filter (Thorlabs FEL0750). This beam is then focused in the horizontal direction through a cylindrical lens CL onto the linear 32x1 Single Photon Avalanche Diode (SPAD) array \cite{Zappa2007,Guerrieri2010a}. Both photon detectors then signal the time tagging FPGA device (UQdevices).}
\label{fig:Setup}
\end{figure} 

The experimental setup depicted in \figref{fig:Setup} is based on a SPDC source using a 1 mm long $\beta$-Barium Borate (BBO) cut at $42^\circ$ for type II frequency degenerate phase matching at $\lambda_0=808$ nm with opening angle $\alpha_s = \alpha_i = 3^\circ$ when the pump is set for normal incidence $\alpha_p = 0^\circ $.  A pump beam is focused in the nonlinear crystal and we alter its propagation direction angle $\alpha_p$ at the crystal by rotating a 1 cm thick piece of NBK7 glass placed in the beam path. The pump beam actual direction $\alpha_p$ is analyzed using a system consisting of a lens (L2) placed at the focal distance away from the BBO crystal and a CCD camera.

An idler photon is  collimated and spectrally filtered by passing through a $10$ nm FWHM bandpass filter centred at $808$ nm which is incorporated in our model by function $\Lambda(\omega_i)$. It is then coupled into a fiber and finally it is detected by a single photon counting module (SPCM). The fiber is SMF-28e+, which supports seven modes at $808$ nm. 

In turn, a signal photon is collimated upon exiting the BBO crystal, passes through a longpass filter (LP) and then is focused horizontally to match the size of the single photon avalanche diode (SPAD) array. We use 32 x 1 pixel SPAD array \cite{Guerrieri2010a}, which  allows us to determine the arrival position of the signal photons and correlate them with detection of the idler photon by SPCM. The array consists of $20 \mu$m diameter detectors spaced $100 \mu$m from each other. Each detector monitors a direction related to angle $\alpha_s$ which is related to a certain transverse wave vector $x$ component $k_{sx}= 2\pi   \sin(\alpha_s) / \lambda_0$.  The finite size, $d$, of each of the SPADs results in monitoring a certain range of directions related to $\delta k= 2 \pi  d / \lambda_0 f$, see \eqref{eq:prob:SPAD}. This allows us to determine the central output angle which the signal photons were emitted at given a fixed idler emission. The photon detections from the SPAD array and SPCM are then recorded and analyzed in a FPGA time tagging unit. 

\begin{figure}
\centering
\begin{tabular}{c}
\subfigure[]{\includegraphics[width=0.85\columnwidth]{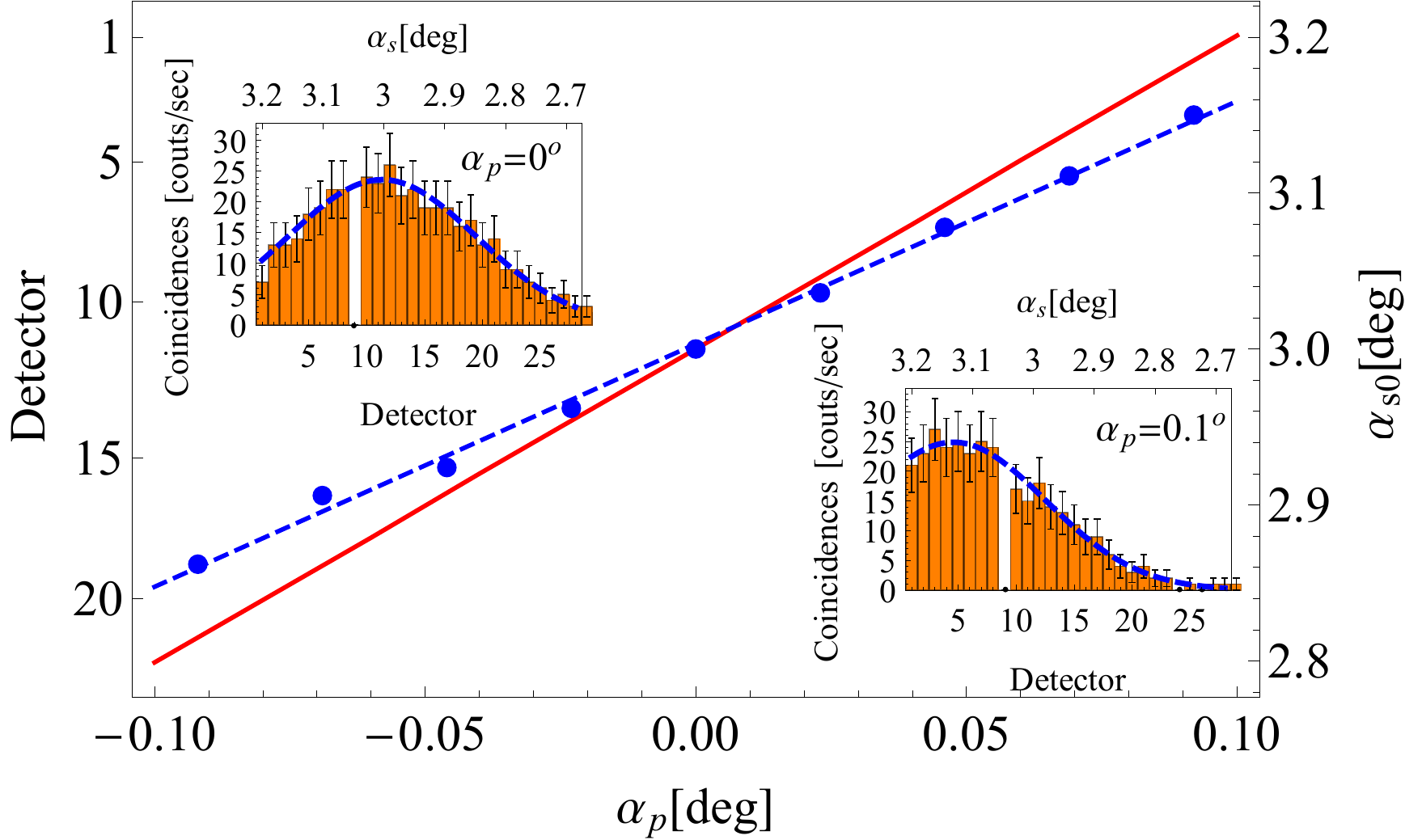}}\\
\subfigure[]{\includegraphics[width=0.85\columnwidth]{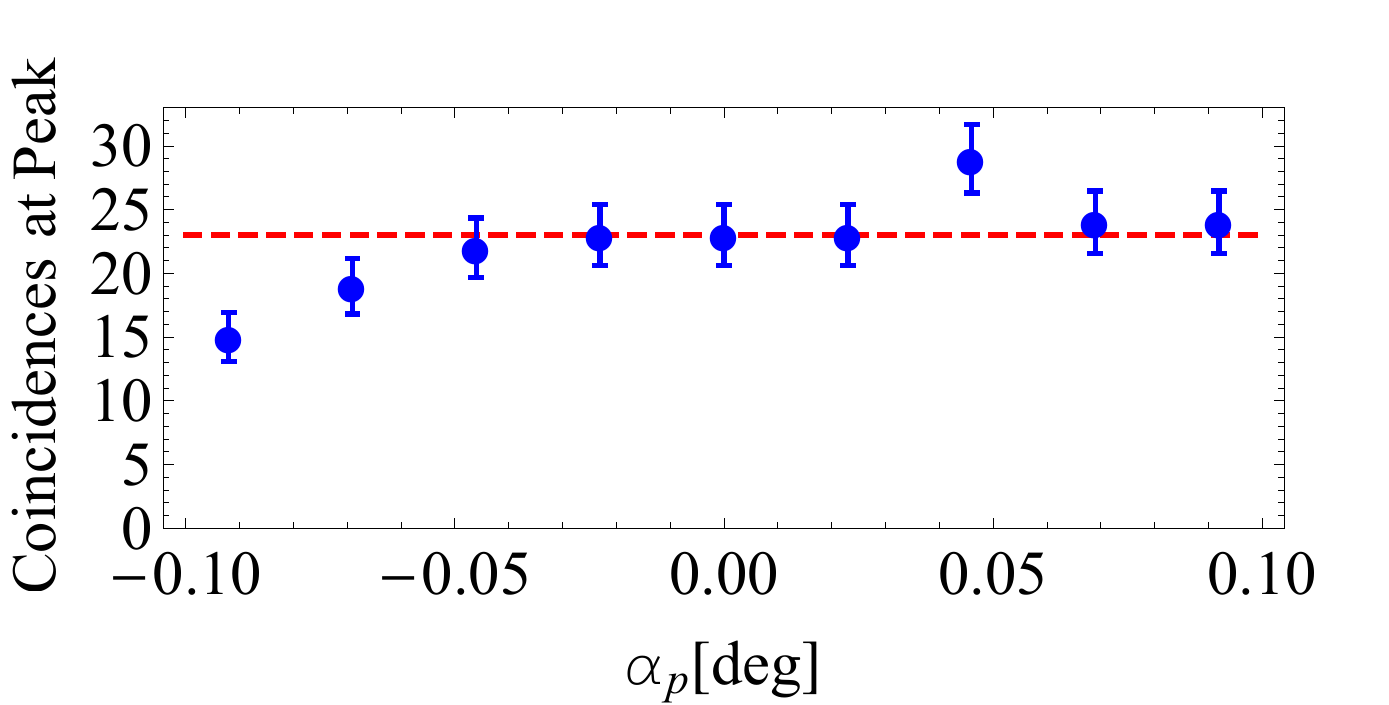}} 
\end{tabular} 
\caption{a) (Inset) Observed counts at the SPAD array, in coincidence with a detection at the SPCM. The data acquisition was set to 2 ns coincidence window and each trial ran for one minute. The pump beam direction was set to (top left) $\alpha_p=0$ and (bottom right) $\alpha_p=0.092^\circ$. The blue solid line shows the best fit with a Gaussian function.  As can be seen in these insets, there is a skipped pixel which was due to very high dark counts on that pixel. The plot shows the maximum from the fit for each set of measurements taken for varying the pump beam central direction $\alpha_p$.  The measured (blue dots) and predicted (red line) signal photon's angle $\alpha_{sc}$ as a function of the input pump angle $\alpha_p$. b)
Maximum count rates at the gaussian centre of the experimental data for various pump angles. The dashed line demonstrates the coincidence count rate at the pump incident angle of $0^\circ$. The count rate drop at values lower than $2.95^\circ$ is an effect of having the cylindrical lens oriented at a slight angle compared to the array. The array is then shifted slightly off the central plane of the signal photons and as the angle continues to shift in this direction the photons are no longer hitting the active area of the SPAD array.
}
\label{fig:angleCorr}
\end{figure}

We measure the photon detection coincidences between the fiber collected idler photon and each of the pixels of the SPAD array detecting the correlated signal photons. In order to maximize the signal to noise ratio we set the coincidence time window at $2$ ns, which is chosen to match the combined effect of detectors' timing jitters and resolution of our electronics. We performed 9 measurements changing the pump beam direction in the range of $-0.1^\circ$ to $0.1^\circ$. The exemplary coincidence statistics for pump beam direction $\alpha_p=0$  and $\alpha_p\approx 0.1$ are inset in \figref{fig:angleCorr}a. Fitting a Gaussian function to the spatial data gives: a) the detector number for which the maximal number of counts were observed, which gives an estimate of the central propagation direction $\alpha_{s0}$ of a signal photon and b) the widths of the beams. The estimated, based on measurements, signal photon propagation direction is depicted using blue markers in \figref{fig:angleCorr}a. The red line presents the theoretical prediction from the model. The blue dashed line represents a linear fit to the experimental data.

It is noted that the theory differs from the experimental observation somewhat as the pump beam input angle, $\alpha_p$, is varied.  The best fit to the experimental data gives: $\alpha_{s0}=3.003(2)+1.56(4)\alpha_p$. The difference of the slopes for the numerical model and experimental data is attributed to the specific distance between the lens and the SPAD array, which is different from the focal length. However, assuming the spatial mode of the propagating photon to be Gaussian, one can predict the width and central propagation direction at each point in the setup.

The measured beam diameters are given in \tabref{tab:widths}. Taking that into account, we concluded that the characteristic width and the central propagation direction are scaled by the same scaling factor, which depends on the position of the lens L3, \eqref{eq:scaling}. Based on the simulated and measured widths of the beams we estimate the scaling factor to be $1.3(1)$. Using this scaling factor our estimated slope is $1.3(1)*1.56(4)=2.2(3)$. This agrees with the theoretical prediction within the statistical error. 

\begin{table}
\small
\npdecimalsign{.}
\nprounddigits{2}
\begin{tabular}{c ||n{3}{2} n{3}{2} n{3}{2} n{3}{2} n{3}{2} n{3}{2} n{3}{2} n{3}{2} n{3}{2}}
\hline
$\alpha_p$ & -0.092 & -0.069 & -0.046 & -0.023 & 0 & 0.023 & 0.046 & 0.069 & 0.092  \\
\hline
$2 w_m$ & 18.7427 & 17.7429 & 18.5779 & 20.7219 & 19.1863 & 19.532	& 19.1033	& 21.0486 &	19.3072 \\
\hline
\end{tabular}
\npnoround
\caption{The widths of the spatial mode of the signal photon  measured on the SPAD array, $2 w_m$. A Gaussian intensity profile was assumed and the beam diameters are given at $13.5 \%$ of maximal intensity. The widths are expressed in the number of detectors. The distance between two neighboring detectors is $100\mu m$. }
\label{tab:widths}
\end{table}

The coincidence count rates should remain constant for small variations in the pump although their position will change. Fig. \ref{fig:angleCorr}b  shows the coincidence  count rate observed at the central position on the array as a function of pump input angle to be almost constant, which is expected from our theoretical model. There is a slight drop as the angle decreases below $-0.05^\circ$ which can be attributed to the cylindrical lens having a slight angle relative to the orientation of the SPAD array and the distance between the SPAD array and lens L3, which differed from the focal length. The result of this angular difference is the photons falling off the active area of the SPAD cells as the angle is shifted towards one direction in the pump. 

\section{Discussion} \label{sec:discussion}

In this work we presented a theoretical model demonstrating the correlations between photons created through the process of SPDC and the pump photon. This model takes into account the spatial correlations when the input angle of the pump beam was manipulated in one spatial dimension and the effect on the down converted photons was calculated. We were able to implement an experiment to measure the spatial position of the down converted photons as a function of the input angle of the pump laser in a SPDC experiment, using a time-resolved SPAD array. Though the theory and experiment differed, the data trend is clearly visible and possible reasons for this difference were also discussed. The results presented here can also be useful for the implementation of multidimensional quantum states encoded in spatial mode \cite{Neves2009} of a single photon and can be used to minimize atmospheric beam wander effects on photon transmission through the atmosphere \cite{Minozzi2013,Mair2001}.

\section*{Acknowledgements}
The authors  acknowledge  funding from NSERC (Discovery, CGS), Ontario Ministry of Research and Innovation (ERA program), CIFAR, Industry Canada and the CFI.  PK acknowledges support by Foundation for Polish Science under Homing Plus no.~2013-7/9 program supported by European Union under PO IG project, and by the National Laboratory FAMO in Torun, Poland and by Polish Ministry of Science and higher Education under Iuventus Plus grant no.~IP2014 020873. The authors would also like to thank Simone Tisa and Franco Zappa for their contribution in developing the detectors array.

\appendix

\end{document}